\documentclass[twocolumn,showpacs,superscriptaddress,article,floatfix]{revtex4}
\usepackage{graphicx}
\usepackage{amsmath}
\usepackage{natbib}
\usepackage{bm}
\usepackage[dvips]{color}

\begin{document}
\title{Quantum Size Effects in Pb/Si(111) Thin Films from Density Functional Calculations}
\date{\today}
\author{M. Rafiee}
\author{S. Jalali Asadabadi\footnote{E-mail: sjalali@phys.ui.ac.ir; Tel: +98-0311-7934176; Fax: +98-0311-7932409}}
\affiliation{Department of Physics, Faculty of Science, University
of Isfahan (UI), Hezar Gerib Avenue, Isfahan 81744, Iran}

\begin{abstract}
The Pb/Si(111) thin films were simulated within the density
functional theory (DFT). The well-known Perdew-Burke-Ernzerhof (PBE)
version of the generalized gradient approximation (GGA) and its
recent nonempirical successor Wu-Cohen (WC) issue were used to
estimate the exchange-correlation functional. Lattice parameters
were calculated for Bulk of the Pb and Si compounds to obtain more
reliable lattice mismatch at the interface to be consistent with our
used full-potential method of calculations. The WC-GGA result
predicts the lattice constants of the Pb and Si compounds better
than the GGA when compared with experiment. We have found that the
spin-orbit coupling (SOC) does not significantly influence the
results. Our finding is in agreement with the recent observation of
the Rashba-type spin-orbit splitting of quantum well states in
ultrathin Pb/Si(111) films. Our result shows, in agreement with
experiment, that the top site (T1) is the most stable phase. A
combination of tight $\sigma$ and feeble $\pi$ bonds has been found
at the interface, which results in a covalent Pb-Si bond. Our
calculated electric field gradient (EFG) predicts quantum size
effects (QSE) with respect to the number of deposited Pb layers on
the Si substrate. The QSE prediction shows that the EFG dramatically
drops on going from first to second layer. The EFG calculation shows
that this system is not an ideal paradigm to freestanding films.

\end{abstract}

\maketitle

\section{Introduction}
\label{sec-intro} The Pb/Si(111) thin films can be considered as an
ideal interface, since it is an unreactive metal-semiconductor
system which forms perfectly an abrupt interface without
intermixing.\cite{Lay91} Intensive experimental\cite{Wei92, Sve08}
and theoretical\cite{Cha03, Cha06} works performed on this system
show that it is a subject of intense technological and scientific
interest with a wide range of applications. Several structures were
proposed experimentally for this system.\cite{Lay91, See95} The
structures vary with their circumstances. The variations form a
complicated phase diagram. The stability of the proposed structural
models were studied theoretically employing pseudopotential
calculations.\cite{Cha03} The complex phase diagram shows that there
are 4 structures with a simple (1$\times$1)-unit cell. The
experimentally observed\cite{Sel00} simple unit cell provides a
suitable case to perform accurate atomic relaxations within an
 all-electron calculation reliably. Our full-potential surface
calculations were then devoted to study these (1$\times$1)
structures with the concentration on their most stable phase.
Quantum size effects (QSEs) in nanostructures, e.g., a
two-dimensional (2D) ultrathin metal films, appear as dramatic
oscillations in many physical properties upon variation of film
thickness.\cite{Tri07, Vaz09} The motion of electrons in the film
plane of such a 2D-system is essentially free, whereas the electrons
are confined in the normal direction to the film surface, which
leads to the quantized electronic states, i.e., quantum well states
(QWS).\cite{Jia07} The oscillation period was found to be
7-monolayer-height-islands of lead grown on top of a silicon
substrate for the electron intensity curve as a function of the
normal component of the electron momentum transfer.\cite{Tri07}
V\'{a}zquez de Parga and coworkers observed the QSEs for a number of
surface properties, i.e., the surface roughening temperature, work
function, chemical reactivity, or the surface diffusion barrier, in
the Pb/Si(111) and Pb/Cu(111) thin films by Scanning Tunneling
Microscopy/Spectroscopy (STM/STS).\cite{Vaz09} The step height of
layer N is related to the electron spillage length into the vacuum
which depends on the DOS at the Fermi level.\cite{Vaz09} Quantum
size effects (QSEs) were found for the total energies and energy
differences in freestanding Pb(111) thin films by using
pseudopotential method as embodied in the CASTEP code.\cite{Mat01}
Influence of the Cu(111) substrate on the geometry structure of the
Pb(111) layers was also indirectly taken into account by solely
reducing in-plane lattice constant to impose 3.3\% lateral
compression on all the Pb(111) layers to mimic the role of Cu(111)
substrate.\cite{Mat01} The effects of Cu(111) substrate were found
to be substantial, as the step heights were calculated to be in more
agreement with experiment within the later in-pane
strain.\cite{Mat01} Wei and Chou theoretically, employing ultrasoft
pseudopotential calculations, studied the quantum size effects
(QSEs) in the clean surface of Pb(111).\cite{Wei02} Recently P. S.
Kirchmann et al. experimentally observed quantum size effects (QSEs)
in the Pb/Si(111).\cite{Kir07} They\cite{Kir07} found their
experimental results for the actual Pb/Si(111) slab to be in
agreement with the pseudopotential results\cite{Wei02} for the
hypothetical freestanding Pb(111) slab. Dil et al.\cite{Dil07}
comparing their Pb/C full-potential calculations with the Pb/Si(111)
experimental results of the others\cite{Upt04} reported that unlike
for Pb on graphite, the Pb overlayer lattice structure is influenced
by that of the Si(111). Here we then intend to, going beyond the
hypothetical freestanding Pb(111) approximation, quantitatively
assess the effects of electronic and crystalline structures of the
underlying Si substrate on the deposited Pb films. For this reason
we inspect the electronic structures at the interface of the
Pb/Si(111) slab. From our electronic structure calculations, we show
that there is a mixed state composed of weak $\pi$ and strong
$\sigma$ bonds between Pb and Si at the interface which results in a
strong Pb-Si covalent bond. We aim to investigate whether the
substrate can influence the results. In order to accomplish the
investigation, we have calculated the work function, energy
differences and surface formation energy as a function of number of
Pb layers for the $(1\times1)$-Pb/Si(111). Our result shows that the
effect of Si substrate can be of significant importance for some
physical quantities depending on their sensitivity to the valence
electron charge density.
 We have observed the quantum size effects
(QSEs) for the $(1\times1)$-Pb/Si(111) thin films. The goal of this
work more transparently is achieved by presenting a physical
interpretation for our QSE calculations for this system in the
electric field gradient (EFG) as an extremely sensitive quantity to
the valence electron charge density distribution. We would also
examine the effect of spin-orbit coupling (SOC) on our thin films.
The result, in agreement with experiment,\cite{Hug08} shows that the
SOC has a minor effect for the Pb/Si(111) thin films. Low
temperature $\sqrt3\times\sqrt3$, 3$\times$3 and
$\sqrt7\times\sqrt3$ phases of Pb/Si(111) were theoretically studied
employing pseudopotential method.\cite{Cud08} It was shown that
there could be a discrepancy in predicting the ground state of
Pb/Si(111) system between local density approximation (LDA) and
generalized gradient approximations (GGA).\cite{Cud08} Another
objective of our density functional theory\cite{Hoh64,Koh65} (DFT)
calculations is to go beyond the LDA and GGA by using the
nonempirical GGA recently proposed by Wu-Cohen (WC) for the
exchange-correlation functional,\cite{Wu06} which is expected to
improve metal surface formation energy and lattice constant
calculations\cite{Wu06, Tra07} compared to the
Perdew-Burke-Ernzerhof GGA (PBE-GGA).\cite{Per96}
\section{Computational Details}
\label{sec-compdetails} State-of-the-art calculations were performed
within the density functional theory\cite{Hoh64,Koh65} (DFT) as
implemented in the WIEN2k code.\cite{Bla01} The augmented plane
waves plus local orbital (APW+lo) method\cite{Sjs00, Mad01} has been
used for solving the Kohn-Sham equations\cite{Koh65} employing the
latest version of the generalized gradient approximation (GGA),
i.e., Wu-Cohen (WC), for the exchange-correlation
functional\cite{Wu06}. We set the Muffin-tin radii to $R_{MT}=2.2$
$bohr$ for the Si and to $R_{MT}=2.5$ $bohr$ for the Pb atoms. The
expansion of the wave functions and charge densities were cut off by
the $R_{MT}K_{max} = 7.5$ and $G_{max} = 14$ parameters,
respectively. The full relaxations were performed
 with the criterion of 1 mRy/bohr on the exerted forces. The relativistic
effects were taken into account by including the spin-orbit coupling
(SOC) in a second variational procedure. A set of
$22\times22\times1$ special k points has been used for integrations
over the Brillouin zone of the 1$\times$1 surface cell.

\section{Bulk Structural Properties} \label{sec-bulk}

\begin{table}[!t]
 \begin{center}
 \caption{Lattice parameter, bulk modulus and pressure derivative of bulk modulus of Si
 (Pb) together with the method of calculations and exchange-correlation
 (XC) functional.
 \label{tab1}}
  \begin{ruledtabular}
   \begin{tabular}{lcccccccc}
      Method&XC-Potential&&$a$(bohr)&& $B$(Gpa) && $B'$ &  \\
      \hline
      APW+lo&GGA-WC            && 10.28 (9.31)&& 93 (47.3)&& 4.23 (2.9)&\\
      APW+lo&GGA-PBE           && 10.34 (9.49)&& 88 (42.2)&& 4.16 (5.0)&\\
      APW+lo&LDA               && 10.22 (9.21)&& 96 (53.1)&& 4.37 (3.7)&\\
      PPW&LDA$^{a (b)}$        && 10.17 (9.16)&& 87 (52.6)&& 3.82 (5.2)&\\
      PPW&GGA$^{c (b)}$        && 10.32 (9.56)&& 87 (39.3)&& ----- (4.7)&\\
      Exp.$^{d (e)}$&          && 10.26 (9.36)&& 98 (43.2)&& 4.02(4.9) &\\
   \end{tabular}
  \end{ruledtabular}
 \end{center}
\begin{flushleft}
$^aReference~\textrm{\onlinecite{Cam88}}.$\\$^bReference~\textrm{\onlinecite{Yu04}}.$\\
$^cReference~\textrm{\onlinecite{Pal99}}.$\\$^dReference~\textrm{\onlinecite{Bea70}}.$\\
$^eReference~\textrm{\onlinecite{Voh90}}.$\\
\end{flushleft}
\end{table}

Before going through the interface and surface calculations in
detail, it seems advisable to first shortly report on the bulk
structural properties. In this case, the accuracy of our
calculations could be first assessed, and second more accurate
lattice parameters would be also obtained to simulate the substrate
with an actual and reliable lattice mismatch at the interface.
Furthermore one aims to get more insight into the various
exchange-correlation functionals, specially the recent proposed
WC-GGA\cite{Wu06} potential. The lattice parameters, bulk moduli,
and the pressure derivative of the bulk moduli were calculated for
the bulks of the Si and Pb compounds. The calculations were
performed within the LDA\cite{Kur99}, PBE-GGA\cite{Per96} and
WC-GGA\cite{Wu06} functionals by fitting the total energy as a
function of volume with the Birch equation of state.\cite{Bir78} The
current APW+lo and previous pseudopotential plane wave (PPW)
results\cite{Cam88, Yu04, Pal99} along with the experimental data
\cite{Bea70, Voh90} are presented in Tab.~\ref{tab1}. Our APW+lo
result, compared to the experiment\cite{Bea70, Voh90}, shows that,
for both the Si and Pb cases, the PBE-GGA\cite{Per96} overestimates
the lattice constants and bulk modulii, whereas the LDA\cite{Kur99}
underestimates them. This is consistent with the general features of
the PBE-GGA\cite{Per96} and LDA\cite{Kur99} functionals for solid
states. In particular, as shown in Tab.~\ref{tab1}, our values are
in complete accord with the previous PPW results.\cite{Cam88, Yu04,
Pal99} The WC-GGA\cite{Wu06} predicts the lattice constant of the Pb
compound better than the LDA\cite{Kur99} and GGA regardless of the
APW+lo or PPW method, while our PBE-GGA\cite{Per96} remains superior
for the prediction of its bulk modulus. This can also be the case
for the lattice parameter of the Si compound. From the result
presented in Tab.~\ref{tab1} one can cautiously state that the
WC-GGA\cite{Wu06} functional improves the lattice parameters over
both the LDA\cite{Kur99} and PBE-GGA\cite{Per96}. This statement is
in agreement with the prediction of Wu and Cohen for the most of
their studied compounds\cite{Wu06}. Our results show that lattice
parameters calculated from WC-GGA\cite{Wu06} approximation are very
close to experiment\cite{Bea70, Voh90}. The calculated lattice
parameter of the bulk Si, i.e., 10.28 a.u., will be used in the
forthcoming Sec.~\ref{sec-Slab} for simulating this compound as our
substrate to mimic the effects of the lattice mismatch on the
valence bands at the interface of the Pb/Si(111). Finally, keeping
in mind that the value obtained for the pressure derivative of the
bulk modulus depends on the range of fitting and as a result
quantitative analysis may not be completely reliable, our result
shows that the PBE-GGA\cite{Per96} yields better pressure
derivatives of the bulk moduli for the Pb and Si compounds than
LDA\cite{Kur99} and WC-GGA\cite{Wu06}.

\section{Slab Determination and Optimization} \label{sec-Slab}

\begin{table}[!t]
 \begin{center}
 \caption{The work function, $\phi_{Si}$ (eV), and surface formation energy,
  $E^f_{Si}$ (eV/$bohr^2$), per unit area for the number of Si(111) bilayers
 ($N_{Si}$).
 \label{tab2}}
  \begin{ruledtabular}
   \begin{tabular}{lcccc}
      &$N_{Si}$&$\phi_{Si}$ (eV)&$E^f_{Si}$ (eV/$bohr^2$)&  \\
      \hline
    & 4  & 4.8484 &1.3037& \\
    & 6  & 4.7179 &1.3281& \\
    & 8  & 4.7295 &1.3274& \\
    & 10 & 4.7299 &1.3270&
    \end{tabular}
  \end{ruledtabular}
 \end{center}
\end{table}

\begin{figure}
 \begin{center}
  \includegraphics[width=8.1cm,angle=0]{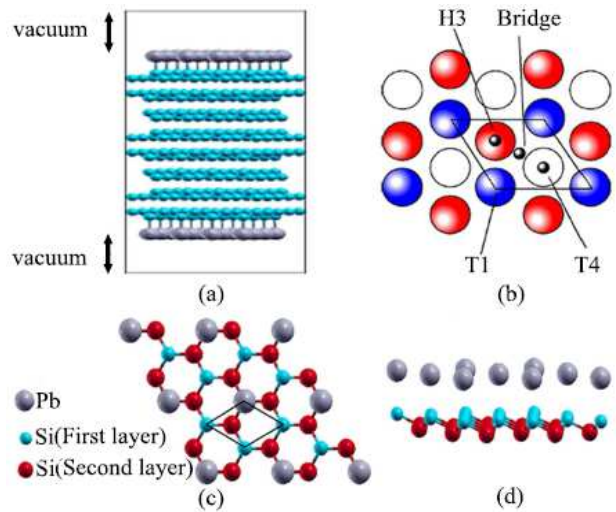}
  \caption{(Color online) (a) A Pb/Si(111) slab symmetrically immersed in a vacuum
   to illustrate atomic arrangement of Pb adatoms (larger grey spheres)
  on 8 bilayers of Si(111) atoms (smaller light spheres).
   (b) A schematic view of Pb overlayers to illustrate T1 (top), H3 (hcp), T4 (fcc)
    and bridge (intermediate point of the T1 and T4 or T1 and H3) sites
   of the Pb adatoms on the Si(111) substrate.
   Top (c) and cross (d) views of the the most stable phase, T1, including a bilayer
    of Si(111) and a Pb overlayer located
   along the first layer of the Si(111) bilayer.
   The (1$\times$1)-unit cell is outlined in (b) and (c). \label{fig1}}
 \end{center}
\end{figure}

The silicon substrate has been simulated using the calculated
WC-GGA\cite{Wu06} lattice parameter in the previous
Sec.~\ref{sec-bulk} along the (111) crystallographic direction of
the $Fd\overline{3}m$ space group. We have found that the Si(111)
substrate, as shown in Fig.~\ref{fig1}(a), can be properly modeled
by stacking 8 Si bilayers, i.e., 16 Si monolayers. Surface formation
energy, $E^f_{Si}$, per unit area and work function, $\phi_{Si}$,
were calculated for the various number of Si(111) bilayers,
$N_{Si}$, and the results are listed in Tab.~\ref{tab2}. Our result
shows that the changes of the work function and the surface
formation energy are less than 0.001 eV by adding more bilayers. The
reliability of the obtained 8 Si bilayers has been further
investigated by comparing the bulk and surface electronic structures
of the silicon. The total densities of states (DOSs) were calculated
for both the bulk and surface of the Si. One expects to observe
similar DOSs for the bulk of Si and the deepest atom from the
surface of the Si(111). Our calculated surface DOS for $N_{Si} < 8$
predicts a wrong metallic behavior for the deepest Si atom. However,
our result confirms that the DOS of the deepest Si layer approaches
to the bulk DOS of the Si semiconductor by approaching $N_{Si}$ to
the 8 bilayers. We avoid to present the DOSs of the substrate
without the deposited Pb layers in this paper, as we just used them
to ensure the validity of the Pb/Si(111) thin films calculations.
The Pb layers were then deposited over and below the prepared
Si(111) substrate to create a symmetric Pb/Si(111) slab. In order to
simulate the Pb/Si(111) thin films, the created Pb/Si(111) slab, as
shown in Fig.~\ref{fig1}(a), is symmetrically immersed in a vacuum
as well. The symmetric thin films causes to add inversion symmetry
which gives rise to speed up the interface relaxation. The vacuum
thickness is defined to be the distance between the top edge of the
slab and the bottom edge of its next neighbor. In order to determine
the vacuum thickness, total energy, work function and exerted forces
on the surface atoms were calculated for the 8 bilayers Si(111)
structure versus various vacuum thicknesses. Our results, which are
not presented here, show that 12 $\AA$ is sufficient for the vacuum
thickness to avoid interactions with the nearest neighbors of the
slab along the Cartesian z axis. The total energy, work function and
forces were well converged in the vicinity of the obtained vacuum
thickness -- they did not show significant changes by increasing the
vacuum thickness more than the value obtained. All the Si underneath
atoms and Pb adatoms are allowed to fully relax by adjusting their
heights, but not their coordinates parallel to the Si(111) surface.
The top (T1), fcc (T4), hcp (H3), and bridge (B2 -- intermediate
point of the T1 and T4 or T1 and H3) sites, as corresponding phases
for the deposition of the Pb layers on the Si(111) surface, are
shown in Fig.~\ref{fig1}(b). In order to better visualize the most
stable phase, top and cross views of the
($\sqrt{3}$$\times$$\sqrt{3}$)-unit cell of the T1 phase are
illustrated in Figs.~\ref{fig1}(c) and (d), respectively. In this
phase, as shown in Fig.~\ref{fig1}(d), the Pb adatoms were
positioned along the first layer of the the first Si(111) bilayers.
The (1$\times$1)-unit cell as the interface structure for our
full-potential all electron ab initio calculations is outlined in
Figs.~\ref{fig1}(b) and (c).

\section{Interface Properties} \label{sec-Interface}
\subsection{Stability} \label{sec-Stability}

\begin{table}[!t]
 \begin{center}
 \caption{Relative total energies with respect to the energy of the T1 phase, E(T1),
  in the presence and absence of the spin-orbit coupling (SOC).
 Here the E(T1) is arbitrarily chosen to be zero.
 \label{tab3}}
  \begin{ruledtabular}
   \begin{tabular}{lcccc}
      SOC&T1 (eV)&T4 (eV)&H3 (eV)&  \\
      \hline
       No     & 0.00  & 0.41 &0.32&\\
       Yes    & 0.00  & 0.25 &0.20&\\
       No$^a$ & 0.00 & 0.35 &0.18&
    \end{tabular}
  \end{ruledtabular}
 \end{center}
 \begin{flushleft}
$^aReference~\textrm{\onlinecite{Bro02}}.$\\
\end{flushleft}
\end{table}

In this section, we would make sure about the most preferable
configuration. Total energies were then calculated for all of the
illustrated structures in the Fig.~\ref{fig1}(b), apart from the
bridge-site structure. The calculations were performed in the
presence and absence of the spin-orbit coupling (SOC). Total
energies relative to the T1 site are presented in Table.~\ref{tab3}.
Our result with and without SOC shows that the T1 site is the most
stable structure among the other configurations. This result is in
agreement with experiment.\cite{See95} The energy, as shown in
Table.~\ref{tab3}, increases from T1 site to T4 site and deceases
from T4 site to H3 site. The later trend, as shown in
Table.~\ref{tab3}, is not affected by the spin-orbit interactions.
The result shows that the effect of spin-orbit interactions, keeping
the trend, is to shift downward total energies of all the sites. Our
results are also quantitatively comparable with the pseudopotential
calculations.\cite{Bro02} The energy of the bridge-site phase has
not been calculated, since, as mentioned in the previous
Sec.~\ref{sec-Slab}, B2 is an intermediate site of T1 and T4 sites
or T1 and H3 sites. Therefore, one can most likely predict that the
surface formation energy of the B2 site because of its symmetry is
somewhere between T1 and T4 sites or between T1 and H3 sites. This
prediction is in complete accord with the pseudopotential
results.\cite{Bro02}
 We will concentrate only on the lowest-energy T1 site from later on
in the remaining subsequent sections.

\subsection{Si-Pb bond length} \label{sec-Bond}

\begin{table}[!t]
 \begin{center}
 \caption{Pb-Si bond length, d(Pb-Si), together with the method of calculations and
  exchange-correlation
 (XC) functional.
 \label{tab4}}
  \begin{ruledtabular}
   \begin{tabular}{lccc}
      Method&XC-Potential &d(Pb-Si)$(\AA)$&  \\
      \hline
        FP-APW+lo&WC-GGA    & 2.72 &\\
        FP-APW+lo&PBE-GGA   & 2.70 & \\
        FP-APW+lo&LDA       & 2.68 & \\
        PPW$^a$  &LDA       & 2.66 & \\
        EXP.$^b$ && $2.66\pm0.03\leq d \leq 2.98\pm0.03$&
    \end{tabular}
  \end{ruledtabular}
 \end{center}
\begin{flushleft}
$^aReference~\textrm{\onlinecite{Cha03}}.$\\
$^bReference~\textrm{\onlinecite{Kum00}}.$\\
\end{flushleft}
\end{table}

Bond length of Pb-Si, d(Pb-Si), at the interface of our slab has
been calculated for the most energetically stable phase, i.e., T1
site, as shown in Tab.~\ref{tab3}. The WC-GGA\cite{Wu06},
PBE-GGA\cite{Per96} and LDA\cite{Kur99} functionals were used for
the exchange-correlation (XC) potential. Our full-potential results
together with the pseudopotential results of the others\cite{Cha03}
as well as experimental data\cite{Kum00} are listed in
Tab.~\ref{tab4}. The experimental data were obtained by surface
X-ray diffraction based on the Pb/Si(111)-$(^{~~3}_{-1}{~~}^{2}_1)$
model\cite{Kum00}. This model contains 7 Pb adatoms in the
$(^{~~3}_{-1}{~~}^{2}_1)$-unit cell\cite{Kum00}. Here we are only
interested in the four of these lead atoms, since they are located
in the top (T1) site as our predicted most stable phase\cite{Kum00}.
The Pb-Si bond lengths for these 4 lead adatoms were
measured\cite{Kum00} to be between $2.66\pm0.03 \AA$ and
$2.98\pm0.03 \AA$.\cite{Kum00} The covalent radii of silicon and
lead are $R_{co}(Si) = 1.18 \AA$ and $R_{co}(Pb) = 1.47 \AA$,
respectively, and the bulk radius of lead is $R_{bu}(Pb) = 1.75
\AA$.\cite{Kum00} Thereby the covalent bond lengths of Pb-Si have to
be between $R_{co}(Si) + R_{co}(Pb)= 2.65 \AA$ and $R_{co}(Si) +
R_{bu}(Pb)= 2.93 \AA$, viz. $2.65 \AA < d_{co}(Pb-Si) < 2.93
\AA$.\cite{Kum00} Our LDA full-potential result, as shown in
Tab.~\ref{tab4}, is in excellent agreement with the LDA
pseudopotential result of the others.\cite{Cha03} The LDA results,
as shown in Tab.~\ref{tab4}, are very close to the strongest
covalent bond length, i.e., $2.65 \AA$. The result demonstrates that
the PBE-GGA and WC-GGA give larger bond lengths which result in
weaker Pb-Si bonds than LDA. This confirms a trend shown in
Tab.~\ref{tab1} in which the LDA underestimates but GGA
overestimates lattice parameters compared to experiment. The weaker
PBE- and WC-GGA bonds, as shown in Tab.~\ref{tab4}, are still in the
covalent bond length interval, i.e. [2.65, 2.93 $\AA$]. The
predicted covalent bonds by LDA and GGA show that the competition
between the Pb-Si and Pb-Pb interactions gives rise to saturate all
the dangling bonds of the silicon substrate. Finally, it worths to
mention that the trend seen in Tab.~\ref{tab1} is reversed here, as
shown in Tab.~\ref{tab4}, between PBE-GGA and WC-GGA in which
PBE-GGA predicts lattice parameter larger than WC-GGA. In contrast
to the lattice parameter trend, here, as shown in Tab.~\ref{tab4},
the Pb-Si bond length within the WC-GGA is slightly larger than the
Pb-Si bond length within the PBE-GGA. Normal and in-plane Pb-Si
distances, viz. $d_z$ and $d_{xy}$, were calculated employing
pseudopotential method for the $\sqrt7\times\sqrt3$ mixed phase
obtained from coadsorption of Pb and Sn on the Si(111) surface. The
Pb-Si bond lengths, d = $({d_{xy}^2+d_z^2})^{\frac{1}{2}}$, were
found to vary from 2.77 \AA~ to 3.33 \AA~ depending on the positions
of silicon atoms with respect to the lead adatoms for the (Pb,
Sn)/Si(111) $\sqrt7\times\sqrt3$ phase.\cite{Cud08} The Pb-Si bond
length of the later $\sqrt7\times\sqrt3$ model with less in-plane
Pb-Si distance can be more comparable with the bond length obtained
from our top site phase. The less in-plane distance and its
corresponding normal distance were calculated to be $d_{xy}$ = 1.79
\AA~ and $d_{z}$ = 2.11 \AA, respectively, which resulted in Pb-Si
bond length d = $(1.79^2+2.11^2)^{\frac{1}{2}}$ = 2.77
\AA.\cite{Cud08} The later bond length, 2.77 \AA, which is in the
above mentioned covalent bond length interval, i.e., [2.65, 2.93
$\AA$] and closer to our WC-GGA result, asserts that the Pb-Si forms
a very strong bond.\cite{Cud08}

\subsection{Work function and surface formation energy} \label{sec-Work}

\begin{table}[!t]
 \begin{center}
 \caption{The work function, $\phi$ (eV), and surface formation energy, $E^f$ (eV/$\AA^2$),
  per (1$\times$1)-unit cell for the Pb at the interface (1 ML coverage) of the Pb/Si(111) together
with the method of calculations, kind of exchange-correlation (XC)
functionals and interactions as well as existence or extinction of
the Si(111) substrate. Since most of the presented pseudopotential
results of the others in this table are for the hypothetical
free-standing Pb(111), our FP-APW+lo results without Si(111)
substrate are also included for the comparison.
 \label{tab5}}
  \begin{ruledtabular}
   \begin{tabular}{lcccccc}
      Method&XC-Pot. &SOC&Si(111)&$\phi$ (eV)&$E^f$ (eV/$\AA^2$)&  \\
      \hline
        FP-APW+lo    &WC-GGA    & No  &Yes&4.72&-0.59&\\
        FP-APW+lo    &WC-GGA    & Yes &Yes&4.65&-0.61& \\
        FP-APW+lo$^a$&WC-GGA    & No  &No&4.23&0.02&\\
        PPW$^b$      &LDA       & No  &Yes&-----&-0.47&\\
        PPW$^c$      &PW-GGA    & No  &No&4.07&-----&\\
        PAW-PPW$^d$  &GGA92     & No  &No&3.83&0.03&\\
    \end{tabular}
  \end{ruledtabular}
 \end{center}
\begin{flushleft}
$^a$These free-standing Pb(111) values are given for N=2. For N=1,
the hypothetical free-standing film is so unstable that may no
longer be considered as a physical system even within a
pure theoretical study.\\
$^bReference~\textrm{\onlinecite{Cha03}}.$\\
$^cReference~\textrm{\onlinecite{Jia06}}$ \\
$^dReference~\textrm{\onlinecite{Sun08}};$ The given surface
formation energy in the reference \onlinecite{Sun08}, 0.372 eV, is
converted from eV to eV/$\AA^2$. Hence the value of 0.372 eV is
divided by the area of the $(1\times 1)$-unit cell, 12.804 $\AA^2$.
 \\
\end{flushleft}
\end{table}
We calculated the work function ,$\phi (eV)$,  as the minimum energy
required to liberate an electron from the Fermi level ($E_F$) to a
point with negligible kinetic energy at the center of the vacuum of
the slab. The calculations were performed using the following
formula:
\begin{eqnarray}
\phi=E_{vac} - E_F, \label{equ1}
\end{eqnarray}
where $E_{vac}$ is estimated by the averaged electrostatic Coulomb
potential at the midpoint of the vacuum of the slab and $E_F$ is the
corresponding Fermi energy. The work function is a sensitive
parameter to the surface conditions, since the liberated electron
must move through the surface. The work functions were calculated at
the interface of the slab with and without SOC and the results are
presented in Tab.~\ref{tab5}. The result shows that the SOC causes
to reduce the work function. The pseudopotential results for the
clean Pb(111) surface\cite{Jia06, Sun08} are also given in
 Tab.~\ref{tab5} within two different exchange-correlation functionals.
The comparison shows that the various versions of GGA can result in
slightly different values. Since the pseudopotential work functions
 are available for the hypothetical free-standing
Pb(111), we also added our results without Si(111) substrate in
Tab.~\ref{tab5} for the comparison. Our free-standing work function
is calculated for the second layer of the Pb(111). A single lead
layer without any substrate may not constitute a meaningful stable
physical system. Our Pb(111) work function which is in good
agreement with the pseudopotential result would be compared with the
results of our actual Pb/Si(111) slab to elucidate the role of
Si(111) substrate. An overlook on the Tab.~\ref{tab5} shows that our
calculated work functions taking silicon substrate into account are
in the same order of magnitudes when compared to the free-standing
full- and/or pseudo-potential results. From the later point, it
appears that the Si substrate does not significantly influence the
work functions. The later point confirms the observation of P. S.
Kirchmann et al.,\cite{Kir07}, where they\cite{Kir07} found the
agreement between their experimental QSE observation for the actual
Pb/Si(111) and the pseudopotential results\cite{Wei02} for the
hypothetical freestanding Pb(111).

The surface formation energy per unit area at zero temperature is
defined as:\cite{Pen00}
\begin{eqnarray}
E^f=\frac{1}{2A}(E_{slab}-N_{Si}E_{Si}^{bulk}-N_{Pb}E_{Pb}^{bulk}), \label{equ2}
\end{eqnarray}
where $E_{slab}$ is the total energy of the slab, $N_{Si}$($N_{Pb}$)
and $E_{Si}^{bulk}$($E_{Pb}^{bulk}$) are the number and bulk energy
of the Si (Pb) atoms in the unit cell, respectively. Here A is the area of the
 (1$\times$1)-unit cell.
 The factor $\frac{1}{2}$ is used in the above formula, since the slab has two
surfaces, as shown in Fig.~\ref{fig1}(a), due to the embodied
inversion symmetry. In principle one can use the above equation as
the standard method to calculate the surface formation energy.
 However, it is well known that in practice the surface formation energy employing
  Equ.~\ref{equ2} diverges.\cite{Fio96, Boe98}
In order to overcome the divergence problem of the standard
Equ.~\ref{equ2}, in this paper we have used
 the following formula\cite{Fio96} to calculate the surface formation energy:
\begin{eqnarray}
E^f=\frac{1}{2A}(E_{slab}-N_{Si}\Delta E_{Si}^N-N_{Pb}\Delta E_{Pb}^N), \label{equ3}
\end{eqnarray}
where $\Delta E_{Si}^N = E_{Si}^{N} - E_{Si}^{N-1}$ and
 $\Delta E_{Pb}^N = E_{Pb}^{N} - E_{Pb}^{N-1}$.
 We have already used Equ.~\ref{equ3}, with $N_{Pb}=0$, in the previous
  Sec.~\ref{sec-Slab} to calculate the surface
formation energy of the pure Si(111) substrate. Our calculated
surface formation energies in the presence and absence  of the
spin-orbit coupling (SOC) along with the pseudopotential results for
both of the actual Pb/Si(111) interface\cite{Cha03} and hypothetical
clean Pb(111) surface\cite{Sun08} are given in Tab.~\ref{tab5}. Our
result shows that the effect of SOC is small as well. Spin-orbit
(SO) coupling breaks down the space inversion symmetry by imposing a
preferable direction, which gives rise to spin splitting. The
inversion symmetry can be also broken at an interface or a surface.
The Rashba-effect\cite{Byc84} on metallic surfaces is a phenomenon,
which originates both from spin-orbit coupling and the lack of
inversion symmetry at surfaces.\cite{Bih06} A Rashba-type spin-orbit
splitting with no coverage dependency was found for quantum well
states formed in ultrathin Pb films on Si (111).\cite{Hug08} Hugo
Dil et al.\cite{Hug08} found for the states at 0.15 and 0.4 eV this
splitting to be 14 and 15 meV, respectively. We found the same
splitting in the vicinity of Fermi level for the valence bands. The
SO coupling within this small splitting, as shown in
Tab.~\ref{tab5}, causes to reduce the work function and surface
formation energy by 0.07 eV and 0.02 eV/$\AA^2$, respectively. As
shown in Tab.~\ref{tab5}, the surface formation energy is a negative
value taking substrate into account, i.e., for the actual Pb/Si(111)
case. On the contrary it is a tiny positive value without substrate,
i.e., for the hypothetical freestanding Pb(111) case. The obtained
negative value when compared with the tiny positive value confirms
that the Pb/Si(111) is energetically more favorable to form that
particular structure than the clean Pb(111) surface. The later point
is in complete accord with the pseudopotential results, as can be
seen from Tab.~\ref{tab5}. Therefore, the effect of substrate is
obviously to make the Pb layers energetically stable.

\subsection{Density of states (DOS)} \label{sec-Density}

\begin{figure}[!t]
 \begin{center}
  \includegraphics[width=9cm,angle=0]{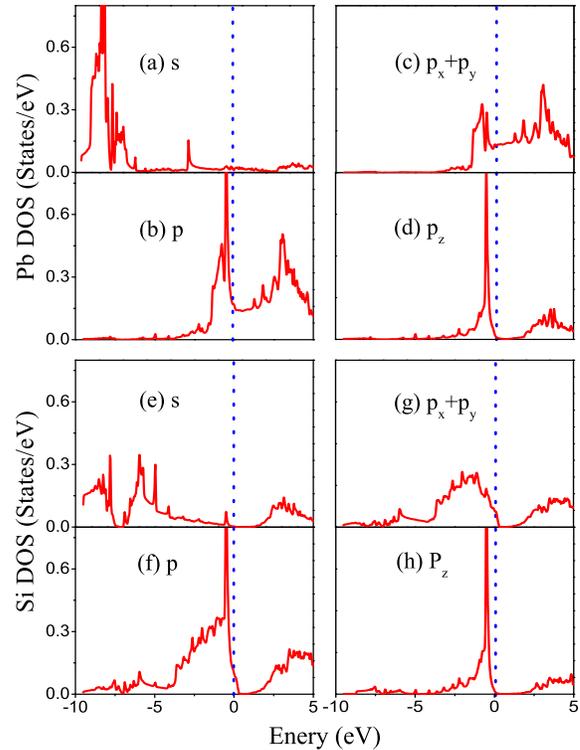}
  \caption{(Color online) Partial densities of states for (a) s Pb, (b) p
  Pb, (c) $p_{x+y}$ Pb, (d) $p_{z}$ Pb, (e) s Si, (f) p Si, (g) $p_{x+y}$ Si,
   (h) $p_{z}$ Si. \label{fig2}}
 \end{center}
\end{figure}

Total and partial densities of states (DOSs) were calculated at the
interface for both of the underneath Si and the overlayer Pb atoms.
The calculations were performed in the absence and presence of the
spin-orbit coupling (SOC) for the slab shown in Fig.~\ref{fig1}(a).
Total DOSs with and without SOC, which are not shown here, were
compared. Despite the large atomic number of Pb, the comparison
shows that the effect of SOC is not very significant on the valence
electronic structure and can be ignored. This is in accordance with
the experiment in which the magnitude of the Rashba-type spin-orbit
splitting were found for this thin films to be so small that can be
detected by the spin integrated angle resolved photoemission (ARPES)
method.\cite{Hug08} The spin splitting originates from all of the
layers and thereby should be calculated by integrating over the
entire layers.\cite{Hug08} The spin splitting is small in Pb/Si(111)
ultrathin films, because the signs of the splittings of the
Pb-vacuum and Si(111)-Pb interfaces are opposite in direction due to
the larger phase shift at the Pb to the vacuum interface than the
phase shift at the Pb to Si(111) interface.\cite{Hug08} Thus the
total splitting which partly cancels each other results in an
ignorable spin-orbit coupling effect in Pb/Si(111). Therefore from
now on we do not include the spin-orbit interactions in all of our
surface calculations. Partial s and p DOSs are shown in
Figs.~\ref{fig2}(a) and (b) for Pb and in Figs.~\ref{fig2}(e) and
(f) for Si atoms at the interface of the slab, respectively. The
result shows that the s states of the Pb and Si atoms are almost
distributed over the energy interval of (-10, -5) eV, while their p
states are nearly in the (-4, $E_F$ = 0) eV energy window. This
demonstrates that the s states are lower in energy than the p states
for both of these Pb and Si atoms. Consequently, the s states of Pb
and Si are more localized than their p states, which are closer to
the Fermi level when compared with the former s sates. Since the p
orbital manifest itself more closely to the Fermi level, the p sates
show more itinerant character than the s sates. Hence, the p
electrons play more important role than the s electrons in the Pb-Si
bonding. Now we can discuss the kinds and qualitatively the
strengths of the Pb-Si bonds, keeping three points in mind: (i) s
(p) electrons are more localized (itinerant) than p (s) electrons
for both of the Pb and the Si atoms, and (ii) below the Fermi level
interval energy (-10, -5) eV of s states do not overlap interval
energy (-4, 0) eV of p states for both of the Pb and Si atoms, as
well as (iii) Pb sits on-top of Si site vertically along z
direction, as shown in the top site configuration in Fig.~\ref{fig1}
(b). The result shows that the 6s states of Pb adatoms are weakly
hybridized with the 3s states of the Si substrate. According to the
above point (ii) the energy window of 3s Si well overlaps the energy
window of 6s Pb, as shown in Figs.~\ref{fig2}(a) and (e). One may
then promptly speculate that the 6s Pb states can be well hybridized
with the 3s Si states. However, according to the point (i) both of
the 6s and 3s states are nearly localized and so do not
substantially contribute in bonding. Thereby the overlap of the 3s
Si and 6p Pb DOSs can only result in an extremely weak $\sigma$-like
bond, which may be due to its weakness neglected. The 6s Pb states
are even much more weakly bonded to the p Si states compared to the
former extremely weak $\sigma$-like Pb-Si bond. According to the
property (ii) energy windows of 6s Pb and 3s Si do not overlap,
because the s-Pb states are over (-10, -5 eV), while the p-Si state
are in (-4, 0 eV), as can be clearly seen from Figs.~\ref{fig2}(a)
and (f). This in turn gives rise to even weaker bond between 6s Pb
and p Si states than the last
 6s Pb and 3s Si bond. This is also the case, as shown
in Figs.~\ref{fig2}(b) and (e), for the p Pb and s Si DOSs. Till
now, there are no any evidence of the appearance of Pb-Si bonding.
Let us then discuss the bonding states by concentrating solely on
the most important p states of the Pb and Si. First, the result
shows that the energy window of p Pb well overlaps the energy window
of p Si, as shown in Figs.~\ref{fig2}(b) and (f). Second, according
to the above mentioned (i) point, the p Pb and Si electrons are more
itinerant than their s electrons. Third, there are a lot of p Pb and
p Si states at an energy very close to the Fermi level, see two
sharp peaks in Figs.~\ref{fig2}(b) and (f). Therefore, the p states
of the Pb and Si atoms can be well hybridized with each other. In
order to determine the kind of Pb-Si bond, the p states of Pb and Si
were further decomposed, as shown in the right panel of the
Figs.~\ref{fig2}, to their partial $p_z$ and $p_{xy}$ states. The
distributions of the $p_z$ and $p_{xy}$ DOSs of both Pb and Si atoms
show that they are completely hybridized with each other. Here, the
z axis is taken to be perpendicular on the surface and x,y are
parallel to the surface. The Pb overlayer atom is just located above
the Si underneath atom along the z-Cartesian coordinate in the top
site configuration, as shown in Figs.~\ref{fig1}(b). Hence, one lobe
of the $p_z$ orbital of Si underneath atom overlaps with one lobe of
the $p_z$ orbital of Pb overlayer atom. Consequently, hybridization
of $p_z$ orbitals of Pb and Si results in a $\sigma$-bond. The
strength of the $p_z$ Pb and $p_z$ Si $\sigma$-bond can be estimated
by observing that the two sharp peaks at an almost single energy
nearby the Fermi level originate from the $p_z$ Pb and $p_z$ Si
DOSs, as shown in Figs.~\ref{fig2}(d) and (h). This can be taken as
an indication to the fact that there is a tightly $\sigma$-bond
between the the $p_z$ Pb and $p_z$ Si at the interface. This result
is in excellent agreement with the pseudopotential result of the
others\cite{Cud08} for Pb/Si(111) $\sqrt3\times\sqrt7$ phase, in
which a strong hybridization between the $P_z$ Pb orbital and the
$P_z$ dangeling bonds of the substrate had been reported. As shown
in Figs.~\ref{fig2}(c) and (g), two lobes of the $p_{xy}$ Pb overlap
with two lobes of the $p_{xy}$ Si as well, which gives rise to a
$\pi$-bond. The strength of the $\pi$-bond cannot be so large,
because the $p_{xy}$ Pb and $p_{xy}$ Si DOSs are broadened, as shown
in Figs.~\ref{fig2}(d) and (h), in contrast to the sharp narrow
$p_{z}$ Pb and Si DOSs. Thereby, a feeble $\pi$-bond is constituted
by the the $p_{xy}$ Pb and $p_{xy}$ Si at the interface. We conclude
this discussion by stating the following points. There is not a pure
$\pi$ or $\sigma$ state in the Pb-Si bond. Instead there is a mixed
state composed of $\pi$ and $\sigma$ states in the Pb-Si bond. The
combination of these two feeble $\pi$ and strong $\sigma$ bonds can
result in a strong covalent Pb-Si bond. The later point is
consistent with our last discussion in Sec.~\ref{sec-Bond} regarding
the bond length of Pb-Si. In summary, the Pb overlayer can be
strongly absorbed by the Si substrate.

\section{Thin Films Properties} \label{sec-Thin}
\subsection{Total Energy and Energy Differences} \label{sec-Energy-thin}

\begin{figure}[!t]
 \begin{center}
  \includegraphics[width=9cm,angle=0]{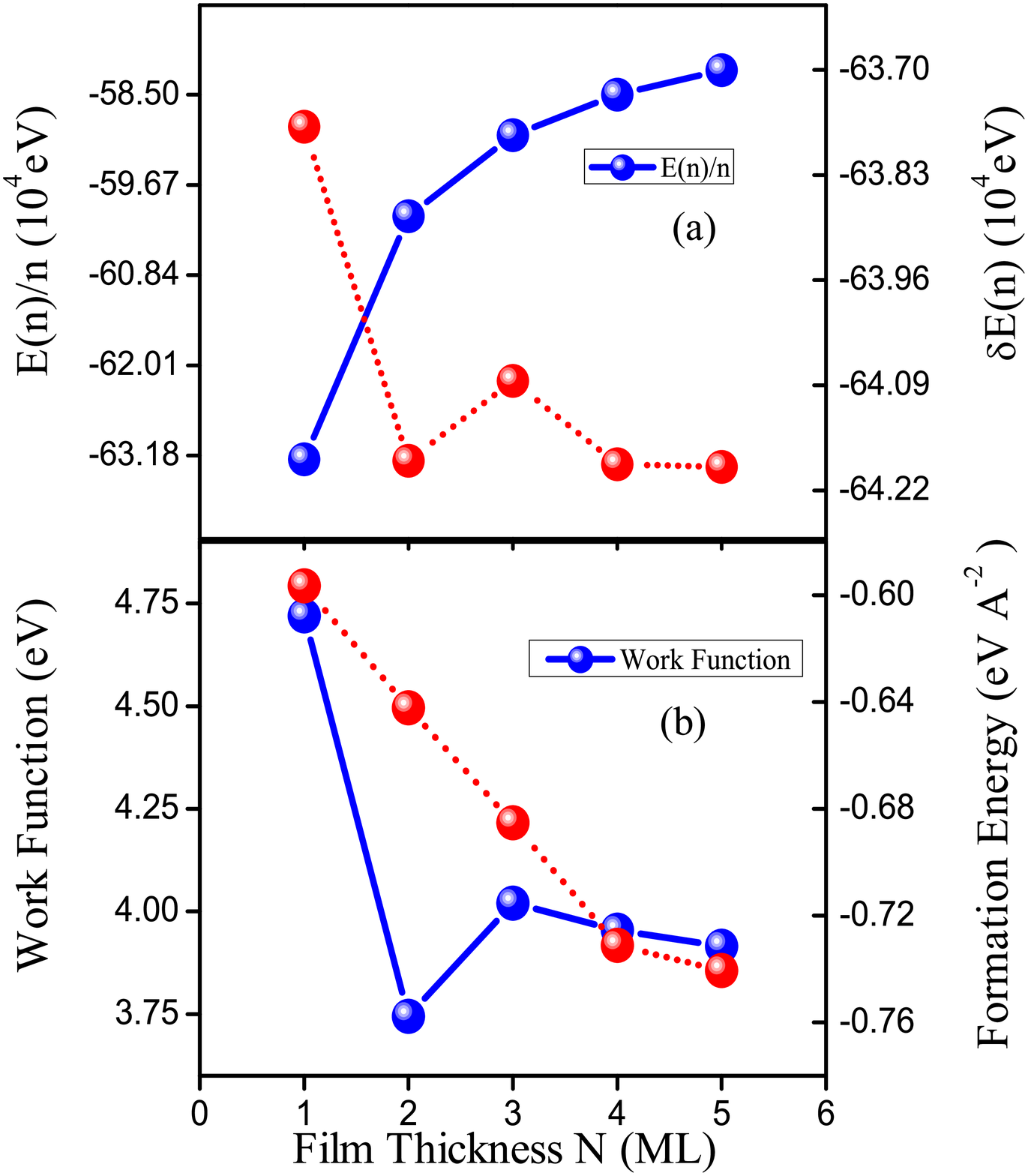}
  \caption{(Color online) (a)Total energy per layer (solid line) and energy
   differences (dotted line) of Pb/Si(111)
  versus number of monolayer (ML) coverages, N.
  (b)Surface formation energy (dotted line) per (1$\times$1)-unit
  cell and work Function (solid line) of Pb/Si(111) versus number of monolayer (ML) coverages, N.
 \label{fig3}}
 \end{center}
\end{figure}

Total energy per number of layers, $E(N)/N$, is shown in
Fig.~\ref{fig3} (a) as a function of the film thickness, N. The
result shows that the $E(N)/N$ increases by N and asymptotically
approaches to a constant value. The asymptotic behavior can be
realized, if we notice that the motion of Pb electrons is confined
in one direction while it remains free to move on the surface. The
later point itself is due to the fact that the Pb/Si(111) is a
prototype of the metal-on-semiconductor system with a high Schottky
barrier against penetrating valence electrons of Pb into the
substrate.\cite{Hes90, Cvi02} The confinement in the nanoscale
causes wave vector k to quantize in a perpendicular direction on the
surface, which results in discrete energy levels associated with the
so-called quantum-well (QW) states. The simplest way to describe
this confinement is then by the one dimensional QW.\cite{Chi00} In a
direction normal on the surface there is no translation symmetry.
The potential in the xy plane is periodic and potential in the z
direction represents the QW. The energy levels in the free electron
approximation are given by:
\begin{eqnarray}
E_{n}=\frac{\hbar^2}{2m}(k_{\parallel}^{2}+k_{\perp}^{2})=\frac{\hbar^2}
{2m}(k_{x}^{2}+k_{y}^{2})+\frac{\hbar^2\pi^2n_{z}^{2}}{2md^{2}},
\end{eqnarray}
where $m$ is the free electron mass and d is the thickness of the
film. The above equation clearly shows that the total energy
approaches to a constant value by increasing the film thickness.
This equation elucidates recognition of the asymptotic behavior
shown in Fig.~\ref{fig3} (a). Monolayer per atom energies
$\frac{E(N)}{N}$ were calculated for the freestanding Pb(111) slab
as a function of monolayers N.\cite{Mat01} In agreement with our
above discussed asymptotic behavior for the actual case of
Pb/Si(111), it was shown that $\frac{E(N)}{N}$ of the hypothetical
freestanding Pb(111) system gradually approaches a constant
value.\cite{Mat01}

The energy difference between two successive layers is defined as:
\begin{eqnarray}
\delta E(N)=E(N)-E(N-1).
\end{eqnarray}
The $\delta E(N)$ versus number of layers, N, is also shown in
Fig.~\ref{fig3}(a). The result shows that the energy difference
oscillates as a function of the number of monolayer (ML) coverages
with the period of $\lambda = 2 ML$. The oscillation is attributed
to the quantum size effect (QSE). Similar QSEs were detected by the
pseudopotential calculations for the freestanding Pb(111) thin
films.\cite{Mat01} One expects that the underlying physics behind of
the observed QSE in the $\delta E(N)$ quantity, as shown in
Fig.~\ref{fig3}(a), can be related to the electronic structure of
the system. We would postpone this point to a more appropriate time
in Sec.~\ref{sec-EFG-thin}, where the electric field gradient (EFG)
as an extremely sensitive quantity to the valence electron charge
density distribution is discussed.
\subsection{Work function and surface formation energy} \label{sec-Work-thin}

\begin{figure}[!t]
 \begin{center}
  \includegraphics[width=9cm,angle=0]{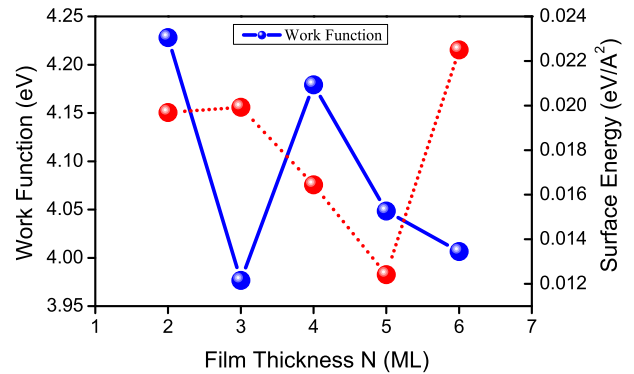}
  \caption{(Color online) Surface formation energy (dotted line) per (1$\times$1)-unit
  cell and work Function (solid line) of freestanding Pb(111) versus number
   of monolayer (ML) coverages, N.  \label{fig4}}
 \end{center}
\end{figure}

The work function, as discussed in the previous Sec.~\ref{sec-Work},
has been calculated utilizing Equ.~\ref{equ1}. The result is
presented in Fig.~\ref{fig3}(b) as a function of Pb thin films
 thickness for the actual Pb/Si(111) slab.
The result shows an oscillatory behavior of the work function versus
the film thickness with the period of $\lambda = 2 ML$. The oscillation is quickly
 damped and the work function is going to be converged
through a few number of Pb layers. The work function oscillation was
previously attributed to the quantum size effects (QSE) and
predicted\cite{Wei02} by using a pseudopotential calculation for the
case of a freestanding Pb slab. The pseudopotential
 result\cite{Wei02} for the hypothetical clean Pb(111) thin films shows that the oscillation is much more slowly damped
 and the work function is converged at a very large number of Pb layers when compared with our full-potential
 results for the real Pb/Si(111) case. The different oscillatory behavior might be related to the effect of the Si substrate.
But one has to notice that the discrepancy  may come from the two different approaches.
 Thus we calculated the work function for the
clean Pb(111) surface within our full-potential method as well. The result as shown
 in Fig.~\ref{fig4} shows that our full-potential work functions are also no longer rapidly damped for the clean Pb(111) case
 in agreement with the pseudopotential result.
This verifies that the oscillatory behavior discrepancy indeed
results mainly from the existence and extinction of the Si substrate
and not from the method of calculations. This authenticates that the
occurred QSE in the work function is influenced by that of Si(111).
The effect of Si substrate on the work function confirms the
calculation of Dil et al.,\cite{Dil07} where they\cite{Dil07}
similarly found that unlike for Pb on graphite, the Pb overlayer
lattice structure is influenced by the Si(111). The effect of our
Si(111) substrate on both of the geometry and electronic structures
of the Pb overlayer atoms is also in agreement with the influence of
the Cu(111) substrate on the geometry structure of the Pb(111)
layers reported by Materzanini and coworkers.\cite{Mat01} In the
later work, the effect of substrate was studied in a tricky manner
without including directly Cu(111) underneath atoms by compressing
the supercell dimensions in the surface plane by 3.3\%.\cite{Mat01}
They found that the explicit inclusion of the Cu(111) substrate
would be highly desirable to explain the remaining quantitative
differences between theory and experiment.\cite{Mat01} It seems that
the later result contradicts our last finding in
Sec.~\ref{sec-Work}. We found in Sec.~\ref{sec-Work} that the order
of magnitude of the work function could not be significantly
affected by the substrate in agreement with the observation
presented in Ref.~\onlinecite{Kir07}. The contradiction can be
resolved if we notice that the work function is not so sensitive
that its order of magnetite, but its oscillatory behavior, can be
affected by the substrate. The electric field gradient (EFG) is much
more sensitive quantity than the work function to the surface
states. We shall show in subsequent Sec.~\ref{sec-EFG-thin} that the
EFG is so sensitive to the substrate that not only its oscillatory
behavior but also its value can be affected by the Si substrate.

We calculated the surface formation energy versus the film
thickness, N, based on the Equ.~\ref{equ2}. The result, which is not
shown here, varies linearly with respect to N. The linear variation
is consistent with the first scheme described in
Ref.~\onlinecite{Fio96}. We then recalculated the surface formation
energy versus N according to the Equ.~\ref{equ3}. The result, as
shown in Fig.~\ref{fig3}(b), manifests a deviation from the linear
behavior on going from fourth to the fifth layer. The deviation is
consistent with the third scheme discussed in
Ref.~\onlinecite{Fio96}. Here we did not use the fourth scheme given
in Ref.~\onlinecite{Fio96}. According to the fourth scheme the Bulk
energies in the standard Equ.~\ref{equ2} can be found by taking the
slope of a fitted straight line to all of the slab total energies
versus N. It is generally believed\cite{Fio96, Boe98, Hon03} that
the divergence problem disappears employing the fourth scheme for
smaller N. However, in this paper we aim to elucidate the role of
substrate. The goal of this paper can be achieved even by the third
scheme which is identical to the the Equ.~\ref{equ3}. Therefore we
calculated the surface formation energy versus N for the clean
Pb(111) surface. The result is shown in Fig~\ref{fig4} which can be
compared with the surface formation energy of the actual Pb/Si(111)
slab as shown in Fig.~\ref{fig3}(b). Furthermore the clean Pb(111)
surface formation energy behaves completely different from that of
its supported slab by the Si substrate. The different behavior
demonstrates the effect of Si on the surface formation energy.

\subsection{Electric Field Gradient (EFG)} \label{sec-EFG-thin}

\begin{figure}[!t]
 \begin{center}
  \includegraphics[width=9cm,angle=0]{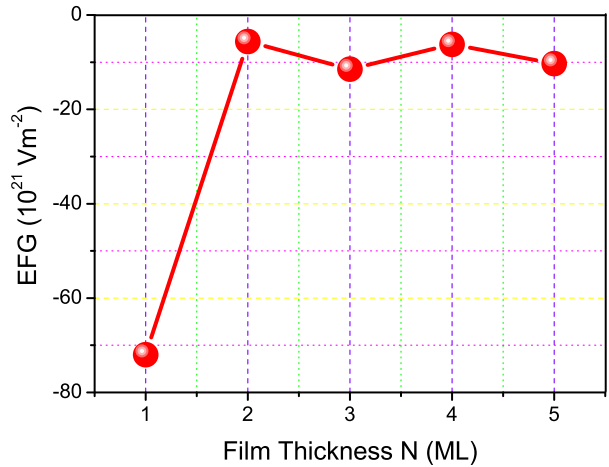}
  \caption{(Color online) The main component of the electric field gradient,
   $V_{zz}$, versus number of monolayer (ML) coverages, N.
 \label{fig5}}
 \end{center}
\end{figure}

\begin{figure}[!t]
 \begin{center}
  \includegraphics[width=8.5cm,angle=0]{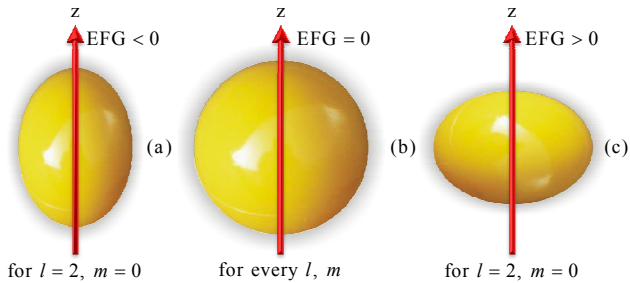}
  \caption{(Color online) A schematic representation of electron charge
   density distribution (ECDD). (a) A longitudinal ECDD deformation
   can yield a nonzero negative value for the main component of the electric
   field gradient, $V_{zz}$, for angular \textit{l}=2 and azimuthal \textit{m}=0 quantum numbers.
  (b) An undeformed spherical ECDD yields
  a zero value for the $V_{zz}$, for every \textit{l} and \textit{m}. (c) A transverse ECDD deformation
  can yield a nonzero positive value for the $V_{zz}$, with \textit{l}=2 and \textit{m}=0.
   \label{fig6}}
 \end{center}
\end{figure}

The electric field gradients (EFGs)\cite{Bla88, Sch90} were
calculated at the Pb sites for our slab. The main component of the
EFG tensor, $V_{zz}$, is presented as a function of film thickness,
N, in Fig.~\ref{fig5}. The sign of the $V_{zz}$ is negative, as
shown in Fig.~\ref{fig5}, for all the layers. The result shows that
for 1 monolayer coverage, i.e., N = 1, the absolute value of the
$V_{zz}$ is a very large number. As shown in Fig.~\ref{fig5}, the
$|V_{zz}|$ is drastically reduced on going from first to the second
layer. Despite the reduction is significant, the $|V_{zz}|$ values
are still considerably large for $N \geq 2$. Our result shows that
for the larger N the $V_{zz}$ oscillates with respect to the number
of monolayer (ML) coverages. The period of oscillation here is also
$\lambda = 2 ML$. Therefore our ab initio calculations clearly
elucidate the quantum size effect (QSE) in the electric field
gradient (EFG). In order to interpret the above observation shown in
Fig.~\ref{fig5} a cartoonlike electron charge distribution is
illustrated in Figs.~\ref{fig6}. The electric field gradient (EFG)
is an extremely sensitive quantity to the deviation from the
spherical valence electron charge density distribution.\cite{Sch90,
Yu91, Jal02, Jal04} The EFG, as shown in Fig.~\ref{fig6} (b), is
zero for a spherical charge density distribution, which is the case
for the cubic or higher point group symmetry. The EFG, as shown in
Figs.~\ref{fig6} (a) and (c), can be nonzero at an atomic site
depending on its point group symmetry for an aspherical charge
density distribution, which is the case for our Pb sites. The EFG
 can be calculated within the APW method by
 $V_{zz}\propto\int[\rho(r)Y_{20}]/r^3dr$.\cite{Las03} The EFG can
 be nonzero for an aspherical charge density distribution and obtained directly from the \textit{l} = 2, \textit{m} = 0
component of the potential expansion inside the muffin-tin
spheres.\cite{Jal04}
  The sign of the EFG depends on the orientation of the
anisotropic charge density distribution. The sign is defined to be
negative (positive) for the charge distribution shown in
Fig.~\ref{fig6} (a) (Fig.~\ref{fig6} (c)). Our surface Pb atoms for
1 monolayer (1 ML) coverage are only supported by their underneath
Si(111) substrate atoms, since for N = 1 there is no more Pb
overlayer on the first deposited Pb layer. The Pb-Pb bonds in the
(x, y)-plane compensate each other. However, there is no Pb-Pb bonds
along the z Cartesian direction to compete with the Pb-Si bonds to
compensate them. The Pb-Si bond, as discussed in
Sec.~\ref{sec-Bond}, is a strong covalent bond. The uncompensated
covalent Pb-Si bond results in a dramatic electron charge density
deformation along the z direction. Consequently one can, according
to the Fig.~\ref{fig6} (a), first expect that the sign of the EFG is
negative and second its value is dramatically large. These
expectations are in excellent agreement with our result as shown in
Fig.~\ref{fig5}. For 2 ML coverages the second Pb layer is supported
by the first Pb layer. The Pb-Pb bonds, which are again compensated
in the (x, y)-plane, are much weaker\cite{Kum00} than the Pb-Si
bonds along the z Cartesian direction. The weaker bonds results in a
less charge density deformation which gives rise to a less value of
the EFG at the Pb site for the second layer. The later point clearly
explains the significant drop of the EFG on going from N = 1 to 2 as
shown in Fig.~\ref{fig5}. Although the EFG is drastically reduced by
adding the second layer, it can be still large due to the anisotropy
of the Pb surface. The anisotropy comes from the fact that for N = 2
there is no Pb-Pb bond over the second layer to compensate the Pb-Pb
bonds between the first and second layers.

Here we would back to the postponed point to clarify that the source
of the quantum size effect (QSE) originates from the electronic
structure of the system. In order to reach the purpose one can first
make a connection between the EFG and the density of states (DOS).
Fortunately an astonishing connection was previously
made\cite{Jal07} between the EFG and the total DOS at the $E_F$,
$N_F$. There it was found an approximately linear relation between
the main component of the electric field gradient and the total
density of states at the Fermi level, viz. $V_{zz} \propto N_F$, in
the $CeIn_3$ compound.\cite{Jal07} Therefore from one side there can
be a relation between the $V_{zz}$ and the $N_F$ in our thin films.
On the other hand here we have shown that the $V_{zz}$ oscillates as
a function of film thickness. Consequently, there might be a
relation between QSE and the electronic structure, or more
specifically, between QSE and the total DOS at the Fermi level
($N_F$). Our finding regarding the later relation between QSE and
$N_F$ verifies the previously reported relation between $N_F$ and
the electron spillage length into the vacuum and as a result between
$N_F$ and the step height of the layer N.\cite{Vaz09} This
illuminates the underlying physics behind of the quantum size
effects (QSEs). Our EFG study shows that the oscillations of a
sensitive physical quantity with respect to the film thickness
originates from the beating of the valence electron charge density
distributions (VECDD). The VECDD deformation can change layer by
layer depending on the bonds strength. The effect of the Si
substrate, which is shown to be of significant importance in the
strength of the bonds, plays an important role in the QSE. One
suspects that the freestanding model can be applicable for those
kinds of the thin films that the strength of their
substrate-overlayer bonds are comparable with the strength of their
overlayer-overlayer bonds. The later point is not the case for our
thin films, since the strengths of the Pb-Si and Pb-Pb bonds are not
comparable.\cite{Kum00} In conclusion our QSE calculation shows that
our system is not an ideal paradigm to freestanding films.

\section{Conclusions}
The effects of Si(111) substrate were investigated on the physical
properties of the Pb/Si(111) thin films. The investigations were
performed within the density functional theory (DFT) employing the
augmented plane waves plus local orbital method (APW+lo). We used
the PBE-GGA and WC-GGA for the exchange-correlation functional. Our
result shows that the WC-GGA is more reliable to study this system.
We included spin-orbit interactions in our the Kohn-Sham
Hamiltonian. Our result shows that the effect of spin-orbit coupling
is not of significant importance for this thin film. Several
structures were considered for the slab. Our ab initio
full-potential calculation shows that the top site (T1) is the most
stable phase. The electronic structures at the interface of the most
stable T1 phase were studied. The study shows that the Pb and Si at
the interface are strongly bonded by $\sigma$ and weakly by $\pi$
bonds. The strong covalent Pb-Si bond is consistent with experiment.
In order to elucidate the role of Si substrate the Quantum size
effects (QSEs) were studied for the T1 configuration. The study is
carried out by calculating the work function and surface formation
energy as well as the electric field gradient (EFG) at the Pb sites.
The calculations were performed as functions of the thin film
thickness (N). We connected the oscillatory behavior of our
calculated physical quantities to the electronic structures at
various layers of the system. The connection has been made by the
concept of the electric field gradient (EFG) as an extremely
sensitive quantity to the electronic structure. The EFG provides a
reliable approach to undertake the underlying physics behind of the
QSE. Therefore the electric hyperfine interaction has merit as a
measurement method to be further used in thin films. Our result
shows that the effect of Si substrate depends on the sensitivity of
the physical quantities to the valence electron charge densities.
The Si substrate can be considerable for the sensitive quantities
such as electric filed gradients (EFG), while it can be ignored for
less sensitive quantities such as the value of the work function.

\acknowledgments This work is supported by University of Isfahan
(UI), Isfahan, Iran. We are also thankful to Computational
Nanotechnology Supercomputing Center Institute for Research in
Fundamental Science (IPM) P.O.Box 19395-5531, Tehran, Iran for the
computing facility.


\begin{thebibliography}{}
\bibitem{Lay91}G. Le Lay, M. Abraham, A. Kahn, K. Hricovini and J. E. Bonnet,
 Physica Scripta. {\bf T35}, 261-267 (1991).
\bibitem{Wei92} H. H. Weitering, D. R. Heslinga, and T. Hibma, Phys. Rev. B {\bf45}, 5991
(1992).
\bibitem{Sve08}  M. \v{S}vec, P. Jel\'{\i}nek, P. Shukrynau, C. Gonz\'{a}lez,
 V. Ch\'{a}b and V. Drchal, Phys. Rev. B {\bf77}, 125104 (2008).
\bibitem{Cha03} Tzu-Liang Chan, C. Z. Wang, M. Hupalo, M. C. Tringides,
 Zhong-Yi Lu, and K. M.
Ho, Phys.Rev. B {\bf68}, 045410 (2003).
\bibitem{Cha06} T.-L. Chan, C. Z. Wang, M. Hupalo, M. C. Tringides, and K. M.
Ho, Phys. Rev. Let. {\bf96}, 226102 (2006).
\bibitem{See95} L. Seehofer, G. Falkenberg, D. Daboul, and R. L. Johnson, Phys. Rev.
B {\bf51}, 13503 (1995).
\bibitem{Sel00} J. Slez\'{a}k, P. Mutombo, V. Ch\'{a}b, Surface Science {\bf454-56},
 584-590 (2000).
\bibitem{Tri07} Michael C. Tringides, Mieczyslaw Jalochowski, and Ernst Bauer, Phys. Today {\bf 60} April 50
(2007).
\bibitem{Vaz09}A.L. V\'{a}zquez de Parga, J.J. Hinarejos, F. Calleja, J. Camarero, R. Otero, R. Miranda, Surface Science {\bf 603}
1389-1396 (2009).
\bibitem{Mat01}Giuliana Materzanini, Peter Saalfrank, and Philip J.D. Lindan, Phys. Rev. B {\bf
63}, 235405 (2001).
\bibitem{Jia07}Jin-Feng Jia, Shao-Chun Li, Yan-Feng Zhang, and
Qi-Kun Xue , J. Phys. Soc. Jpn. {\bf 76}, 082001 (2007).
\bibitem{Wei02}C. M. Wei, and M. Y. Chou, Phys. Rev B {\bf66}, 233408 (2002).
\bibitem{Kir07}P. S. Kirchmann, M. Wolf, J. H. Dil, K. Horn, and U. Bovensiepen,
 Phys. Rev. B {\bf76}, 075406 (2007).
\bibitem{Dil07} J. H. Dil, T. U. Kampen, B. H\"{u}lsen, T. Seyller, and K.
Horn, Phys. Rev. B {\bf75}, 161401(R) (2007).
\bibitem{Upt04} M. H. Upton, C. M.Wei, M.Y. Chou, T. Miller, and T.-C.
Chiang, Phys. Rev. Lett. {\bf93}, 026802 (2004).
\bibitem{Hug08}J. Hugo Dil, Fabian Meier, Jorge Lobo-Checa, Luc Patthey, Gustav Bihlmayer, and J\"{u}rg
Osterwalder, Phys. Rev. Lett. {\bf 101}, 266802 (2008).
\bibitem{Cud08}P. Cudazzo, G. Profeta, A. Continenza, Surface Science {\bf 602}, 747-754 (2008).
\bibitem{Hoh64} P. Hohenberg and W. Kohn, Phys. Rev. {\bf 136}, 864 (1964).
\bibitem{Koh65}W. Kohn and L. J. Sham, Phys. Rev. {\bf 140}, A1133 (1965).
\bibitem{Wu06}Z. Wu, and R. E. Cohen, Phys. Rev. B {\bf73}, 235116 (2006).
\bibitem{Tra07} Fabien Tran, Robert Laskowski, Peter Blaha, and Karlheinz
Schwarz, Phys. Rev. B {\bf75}, 115131 (2007).
 \bibitem{Per96} J. P. Perdew, K. Burke, and M. Ernzerhof,
 Phys. Rev. Lett. {\bf 77}, 3865 (1996).
\bibitem{Bla01}P. Blaha, K. Schwarz, G. K. H. Madsen, D. Kvasnicka, and J. Luitz,
 WIEN2K, "An Augmented Plane Waves + Local Orbitals Program
for Calculating Crystal Properties," Karlheinz Schwarz, Techn.
Universitat Wien, Austria, ISBN 3-9501031-1-2 (2001).
\bibitem{Sjs00} E. Sj\"{o}stedt, L. Nordstr\"{o}m,
 and D. J. Singh, Solid State Commun. {\bf 114}, 15 (2000).
\bibitem{Mad01} G. K. H. Madsen, P. Blaha, K. Schwarz,
 E. Sj\"{o}stedt, and L. Nordstr\"{o}m, Phys. Rev. B {\bf 64}, 195134 (2001).
\bibitem{Kur99} S. Kurth, J. P. Perdew, and P. Blaha, Int. J. Quantum Chem. {\bf 75},
 889 (1999).
\bibitem{Bir78} F. Birch, J. Geophys. Res. {\bf83}, 1257 (1978).
\bibitem{Cam88}P. E. V. Camp, V. E. V. Doren and J. T. Devreese, Phys. Rev. B {\bf38},
 12675 (1988).
\bibitem{Yu04}Dengke Yu and Matthias Scheffler, Phys. Rev. B {\bf70}, 155417 (2004).
\bibitem{Pal99}M. Palummo, G. Onida, R. Del Sole and M. Corradini, and L. Reining,
 Phys. Rev B {\bf60}, 11329 (1999).
\bibitem{Bea70}A. G. Beattie and J. E. Schirber, Phys. Rev. B {\bf1}, 1548 (1970).
\bibitem{Voh90}Y. K. Vohra and A. L. Ruoff, Phys. Rev. B {\bf42}, 8651 (1990).
\bibitem{Bro02}S. Brochard, Emilio Artacho, O. Custance, I. Brihuega, A. M. Baro´,
 J. M. Soler, and J. M. G\'{o}mez-Rodr\'{i}guez,
 Phys. Rev. B {\bf66}, 205403 (2002).
 \bibitem{Kum00}C. Kumpf, O. Bunk, J. H. Zeysing, M. M. Nielsen, M. Nielsen,
  R.L. Johnson and R. Feidenhans'l, Surf. Sci. {\bf448}, L213 (2000).
\bibitem{Pen00}R. Pentcheva, and M. Scheffler, Phys. Rev B {\bf61}, 2211 (2000).
\bibitem{Jia06}Yu Jia, Biao Wu, H. H. Weitering, and Zhenyu Zhang, Phys. Rev B {\bf74},
 035433 (2006).
\bibitem{Sun08}Bo Sun, Ping Zhang, Zhigang Wang, Suqing Duan, Xian-Geng Zhao, Xucun Ma,
 and Qi-Kun Xue, Phys. Rev B {\bf78}, 035421 (2008).
\bibitem{Fio96}Vincenzo Fiorentiniy and M Methfessel, J. Phys.: Condens. Matter {\bf8},
 6525-6529 (1996).
\bibitem{Boe98}J C Boettger, John R Smith, Uwe Birkenheuer, Notker R\"{o}sch,
 S B Trickey, John R Sabin and S Peter Apell,
J. Phys.: Condens. Matter {\bf10}, 893–894 (1998).
\bibitem{Byc84}Y.A. Bychkov and E.I. Rashba, JETP Lett. {\bf 39}, 78 (1984).
\bibitem{Bih06}G. Bihlmayer, Yu.M. Koroteev, P.M. Echenique, E.V. Chulkov, S.
Bl\"{u}gel, Surface Science {\bf 600}, 3888-3891 (2006).
\bibitem{Hes90} D.R. Heslinga, H.H. Weitering, D.P. van der Werf, T.M. Klapwijk,
and T. Hibma, Phys. rev. Let. {\bf 64}, 1589 (1990).
\bibitem{Cvi02} B. Cvikl, D. Koro\v{s}ak, J. Appl. Phys. {\bf 91}, 4281 (2002).
\bibitem{Chi00} T.-C. Chiang, Surf. Sci. Rep. {\bf 39}, 181 (2000).
\bibitem{Hon03}H. Hong, C. M. Wei, M. Y. Chou,  Z. Wu, L. Basile, H. Chen,
 M. Holt, and T.-C. Chiang, Phys Rev. Lett. {\bf90}, 076104 (2003).
\bibitem{Bla88} P. Blaha and K. Schwarz, P. H. Dederichs,
Phys. Rev. B {\bf 37}, 2792 (1988).
\bibitem{Sch90} K. Schwarz, C. Ambrosch-Draxl, P. Blaha,
 Phys. Rev. B {\bf 42}, 2051 (1990).
\bibitem{Yu91} J. Yu, A. J. Freeman, R. Podloucky, P. Herzig, P. Weinberger,
 Phys. Rev. B {\bf 43}, 532 (1991).
\bibitem{Jal02} S. Jalali Asadabadi, S. Cottenier,
 H. Akbarzadeh, R. Saki, and M. Rots, Phys. Rev. B {\bf 66}, 195103 (2002).
\bibitem{Jal04}S. Jalali Asadabadi and H. Akbarzadeh,
 Physica B {\bf 349}, 76-83 (2004).
\bibitem{Las03}R. Laskowski, P. Blaha, and K. Schwarz, Phys. Rev. B {\bf67}, 075102 (2003).
\bibitem{Jal07}S. Jalali Asadabadi, Phys. Rev. B {\bf75}, 205130 (2007).

\end{thebibliography}
\end{document}